\documentclass{PoS}

\title{$O(a^2)$ improvement of the overlap-Dirac operator}

\ShortTitle{$O(a^2)$ improvement of the overlap-Dirac operator}

\author{\speaker{H. Ikeda}%
         \\
        The Graduate University for Studies (Sokendai)\\
        E-mail: \email{hikeda@post.kek.jp}}

\author{S. Hashimoto\\
       High Energy Accelerator Research Organization (KEK)\\
       School of High Energy Accelerator Science, The Graduate University
       for Advanced Studies (Sokendai)\\
       E-mail: \email{shoji.hashimoto@kek.jp}}

\abstract{
  We construct an $O(a^2)$-improved overlap-Dirac operator by designing an
  improved overlap kernel, based on the Symanzik improvement program. 
  Field rotation terms are also identified to improve
  off-shell amplitudes for both massless and massive fermions.
  We check the free dispersion relation and propagator, and show that
  improved results become to close to the continuum ones at low
  momentum region.
  We test the effect of improvement on the full-QCD gauge configuration 
  and find that the relativistic dispersion relation is satisfied
  within a few percent error up to $m_qa \approx 0.5$.
}

\FullConference{The XXVII International Symposium on Lattice Field Theory\\
		 July 26-31, 2009\\
		 Peking University, Beijing, China}

\usepackage{amsmath,amssymb}
\usepackage{makeidx}

\def\qq{\qquad}

\def\slashchar#1{\setbox0=\hbox{$#1$}           
   \dimen0=\wd0                                 
   \setbox1=\hbox{/} \dimen1=\wd1               
   \ifdim\dimen0>\dimen1                        
      \rlap{\hbox to \dimen0{\hfil/\hfil}}      
      #1                                        
   \else                                        
      \rlap{\hbox to \dimen1{\hfil$#1$\hfil}}   
      /                                         
   \fi}                                         %

\def\Dw{D_{\rm w}}
\def\Dov{D_{\rm ov}}

\def\Dsl{\slashchar{D}}

\def\Om{\Omega}
\def\Omb{\bar{\Om}}

\newcommand{\nn}{\nonumber \\}

\let\del=\nabla
\let\De=\Delta
\let\de=\delta
\let\ga=\gamma

\let\ad=\dagger

\def\Omb{{\bar\Om}}

\def\+{\,+\,}

\begin{document}
\section{Introduction}
Discretization effect is one of the most significant sources of the
systematic error in lattice QCD calculations. 
The improvement of lattice action and operators have therefore been
extensively studied since the early days of lattice field theory.
The most well-known and widely used example is the clover fermion
action \cite{Sheikholeslami:1985ij}, which removes the $O(a)$ error in
the Wilson's original lattice fermion action.
According to the Symanzik's improvement program
\cite{Symanzik:1983dc}, it adds a dimension-five operator to the
lattice action to cancel the source of error of $O(a)$ present in the
Wilson fermion action.
A non-perturbative method to tune the parameter in the action has also
been established later \cite{Luscher:1996ug}.
For further improvement, one has to add dimension-six and
dimension-seven operators consecutively, as discussed in
\cite{Alford:1996nx}, for instance.
These highly improved lattice actions are not so popular in the
current lattice QCD simulations, since the action contains many terms
with parameters to be tuned.

One of the reasons for the difficulty of designing highly improved
lattice fermion operator is that the number of operators to be
considered is large because of the explicit violation of the chiral
symmetry in the Wilson fermion action.
Indeed, the $O(a)$ term appears because of the chiral symmetry
violation, while the chirally symmetric lattice actions do not have
this contribution from the beginning as one cannot write down the
relevant operator of dimension-five while preserving chiral symmetry.
The same argument applies at $O(a^{2m+1})$ in general (for $m$ a
positive integer).
In other words, if one starts the improvement program from chirally
symmetric lattice actions, the first error one encounters is $O(a^2)$,
and once it is removed, the next is $O(a^4)$.
Therefore, the effect of improvement is much more dramatic than in the
case of the improvement of the Wilson fermion.
In fact, the $O(a^2)$-improvement of the staggered fermion has been
worked out and used in numerical simulations \cite{Follana:2006rc}.
It uses this property of chirally symmetric lattice fermion action.
When used for heavy quarks, one can greatly accelerate the convergence
to the continuum limit.

In this work we consider the $O(a^2)$-improvement of the overlap
fermion \cite{Neuberger:1997fp}.
The overlap fermion preserves exact chiral symmetry
through the Ginsparg-Wilson relation \cite{Luscher:1998pqa}.
Although the numerical cost is high in the practical use of the
overlap fermion, dynamical fermion simulations have already been
performed by the JLQCD and TWQCD collaborations, from which many
interesting physics results have been obtained thanks to its excellent
chiral property (for a recent summary,
see \cite{Hashimoto:2008fc}).

The improvement can be achieved by two steps, {\it i.e.} improvement
of the action and the field rotation.
Since the form of the overlap fermion is largely restricted by the
Ginsparg-Wilson relation, improvement of the lattice action is done by
modifying the kernel operator to be used to construct the overlap
operator. 
To be explicit, we use the fermion action of Eguchi-Kawamoto
\cite{Eguchi:1983xr} and Hamber-Wu \cite{Hamber:1983qa}, which is
called the D34 action in the convention of \cite{Alford:1996nx}.
Once we remove the Lorentz-violating discretization effects of
$O(a^2)$ by this choice of the kernel operator, remaining errors can
be removed by field rotations.


\section{Formulation of the improved operator}
\label{sec:formulation}

The overlap operator in the massive case is defined by
\begin{equation}
 D_{\rm ov}(m_q) = \left(1 - \frac{am_q}{2\rho}\right) D_{\rm ov} + m_q,
\end{equation}
where the massless operator $D_{\rm ov}$ is given by
\begin{equation}
  D_{\rm ov} 
  = \frac{\rho}{a} \left(1 + \frac{X}{\sqrt{X^\dagger X}}\right)
  \ , \qq X = D_{\rm w} - \frac{\rho}{a}.
\end{equation}
The parameter $\rho$ controls the large negative mass of the overlap
kernel. 
The conventional choice for the kernel operator is that of the Wilson
fermion $D_{\rm w}$, which is
\begin{equation}
 \Dw = \sum_\mu (\gamma_\mu \nabla_\mu - \frac{1}{2}a \Delta_\mu) 
 \sim \slashchar{D} - \frac{a}{2} D^2 + O(a^2).
\end{equation}
Near the continuum limit, it reduces to the continuum Dirac operator
$\slashchar{D}$ plus the $O(a)$ error coming from the Wilson term.
$\nabla_\mu$ and $\Delta_\mu$ are first- and second-order covariant
lattice derivatives, respectively.

Near the continuum limit, the overlap operator with the Wilson kernel
becomes 
\begin{equation}
\label{dov_pert}
 \Dov = \Dsl - \frac{a}{2\rho}\Dsl^2 + \frac{a^2}{6} \sum_\mu \gamma_\mu
 D_\mu^3 + \frac{a^2}{2\rho^2} \left(\Dsl^3 - \frac{\rho}{2} \{\Dsl, D^2 \} \right)+ O(a^3).
\end{equation}
The $O(a)$ term can be simply removed by a field rotation proportional
to $\Dov$, while the $O(a^2)$ terms, especially the third term of
right-hand side which violates the Lorentz symmetry, cannot be removed.
The usual overlap operator thus has an $O(a^2)$ discretization error.
To remove the Lorentz-violating term, we introduce the improved
kernel, which is closer to the continuum limit
$\Dw ' \sim \Dsl + O(a^3)$.
Then, the overlap operator takes a simple form up to $O(a^4)$ errors:
\begin{equation}
  \label{eq:Dov'}
  \Dov ' = \Dsl - \frac{a}{2\rho}\Dsl^2 + \frac{a^2}{2\rho^2}\Dsl^3  -
  \frac{3a^3}{8\rho^3}\Dsl^4 + O(a^4).
\end{equation}
With this operator we can remove the unwanted terms up to and
including the $O(a^3)$ term by a field rotation proportional to
$\Dov$, and the remaining errors start from $O(a^4)$.

As an improved kernel which has no $O(a)$ and $O(a^2)$ errors,
we use the D34 action.
Massless D34 action is defined by
\begin{equation}
 D_{{\rm D34}} = \sum_\mu \del_\mu \left( 1 - b a^2 \De_\mu\right)
 \gamma_\mu \+ \sum_\mu c a^3 \De_\mu^2.
\end{equation}
In order to remove the $O(a^2)$ error at tree level, $b$ = 1/6.
The parameter $c$ is an arbitrary parameter to control the mass of
doublers.
We take $c$ = 1/6 in the following.
For the free case, this action has no $O(a)$ and $O(a^2)$ error, but
it is no longer the case once the gauge interaction is turned on.
In particular, the $O(a)$ term may arise as radiative corrections, and
one has to add another term to cancel it.
The explicit form of this action on the lattice is
\begin{eqnarray}
 aD_{{\rm D34}} & = & 4\de_{x,y} - \frac{2}{3}\sum_\mu
  \left[(1-\ga_\mu) U_{\mu,x} \de_{x+\mu,y} + (1 + \ga_\mu)
   U_{\mu,x-\mu}^\ad \de_{x-\mu,y}\right]\nn
 && + \frac{1}{12}\sum_\mu \left[(2-\ga_\mu) U_{\mu,x} U_{\mu,x+\mu}
                    \de_{x+2\mu,y} + (2+\ga_\mu) U_{\mu,x-\mu}^\ad
                    U_{\mu,x-2\mu}^\ad \de_{x-2\mu,y} \right].
\end{eqnarray}

We now consider the field rotation to remove the remaining
discretization effects.
Starting from the continuum action, 
$\int d^4x \ \bar{\psi}_c(x) (\slashchar{D} + m_q) \psi_c(x)$
with fermion fields $\psi_c$ and $\bar{\psi}_c$, one may define a rotation
\begin{equation}
 \psi_c = \Omega_c \psi \qq \bar{\psi}_c = \bar{\psi} \bar{\Omega}_c,
\end{equation}
which produces the action
$\int d^4x \ \bar{\psi} (x) D_{\rm ov}' (m_q) \psi (x)$
corresponding to (\ref{eq:Dov'}).
Namely, the rotation satisfy the relation
$\Dov ' (m_q) = \bar{\Omega}_c (\Dsl + m_q) \Omega_c$.
So far, the rotation matrices $\Om_c$ and $\Omb_c$ are written in
terms of the continuum operator $\slashchar{D}$.
Note that a field rotation does not affect spectral quantities, as far as
the Jacobian of the transformation is taken into account.
The Jacobian may affect the renormalization of the gauge coupling at the
quantum level but does not matter at the classical level.

There are several choices of the rotations to identify the continuum
Dirac operator as the improved overlap operator up to neglected higher
order terms.
Since the higher powers of the overlap operator, such as $\Dov^2$, in
the lattice action is computationally expensive in practical simulations,
we arrange the field rotation so that they vanish in the lattice action.
Our choice of the field rotation is
\begin{equation} \label{rot}
  \Om_c = 1 - \frac{a}{2\rho}\Dsl + \frac{a^2}{2\rho^2}\Dsl^2 -
  \frac{3a^3}{8\rho^3}\Dsl^3 - \frac{m_q a^2}{4\rho^2} (\Dsl -
 m_q)\left(1 - \frac{a}{2\rho}\Dsl\right), \qq
 \Omb_c = 1 .
\end{equation}
With this choice, the massive improved operator takes a simple form
\begin{equation} \label{Dov_m}
 D_{\rm ov}'(m_q) = \left( 1 - \frac{a}{2\rho}M(m_q,\rho) \right) D_{\rm
 ov}' + M(m_q,\rho) 
\end{equation}
with $M(m_q,\rho) = m_q \left(1 + \frac{m_q^2a^2}{4\rho^2}\right)$.
It means that one can simply use the conventional overlap operator in
the numerical simulation except that the kernel is improved.
Since the rotation operator is proportional to $\Dsl$, the on-shell
quantities are unchanged, and off-shell amplitudes are obtained by
undoing the rotation.
To do so, the lattice version of the rotation is given by
\begin{equation} \label{rot_lat}
 \Omega_L = 1 - \frac{a}{2\rho}D_{\rm ov} '+ \frac{a^2}{4\rho^2}D_{\rm
  ov}'^2 + \frac{a^3}{8\rho^3}D_{\rm ov}'^3 -
  \frac{m_q a^2}{4\rho^2}(D_{\rm ov}'-m_q) - \frac{m_q^2 a^3}{8\rho^3}D_{\rm
  ov}', \qq
  \Omb_L = 1,
\end{equation}
where $\Om_L$ and $\Omb_L$ are the same as $\Om_c$ and $\Omb_c$ up to
the $O(a^3)$ terms.
The off-shell improved propagator is then constructed as
$\Om_L \Dov^{'-1}(m_q) \Omb_L = (\Dsl + m_q)^{-1} + O(a^4)$
\footnote{
  We note that this construction of the rotation has an apparent problem
  that the manifest chiral symmetry of the form
  $\gamma_5 S_F(x,y)+S_F(x,y)\gamma_5=0$ is lost.
  We will discuss on a modification of the lattice action to satisfy
  this condition in future publications.
}, 
which does not require additional inversion of the overlap operator.

\section{Relations at the tree level}
\begin{figure}[tbp]
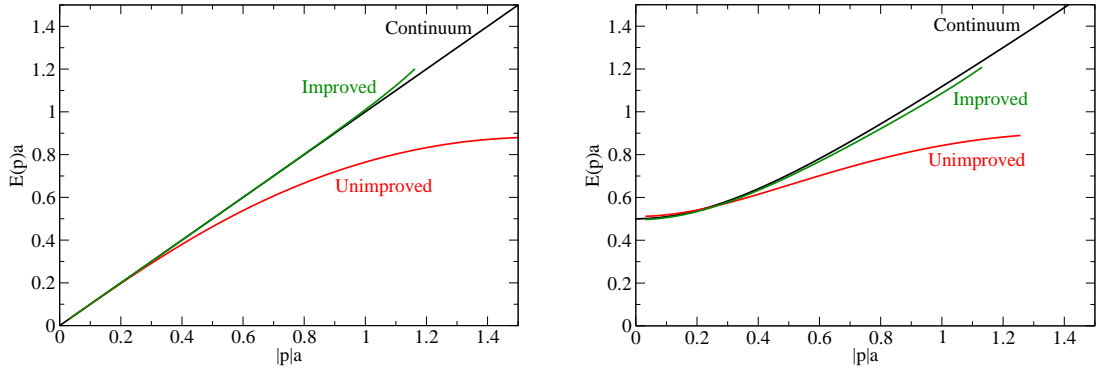

 \begin{tabular}{cc}
  \begin{minipage}{0.5\hsize}
   \includegraphics[width=0.9\textwidth,clip]{disp_m0.eps}
  \end{minipage}
  \begin{minipage}{0.5\hsize}
   \includegraphics[width=0.9\textwidth,clip]{disp_m.eps}
  \end{minipage}
 \end{tabular}
 \caption{Dispersion relation with the Wilson (unimproved) and with
   the improved kernels.
   Left shows the massless case, while the right is at $m_qa=0.5$.
   }
   \label{fig:dispersion}
\end{figure}

\begin{figure}[tbp]
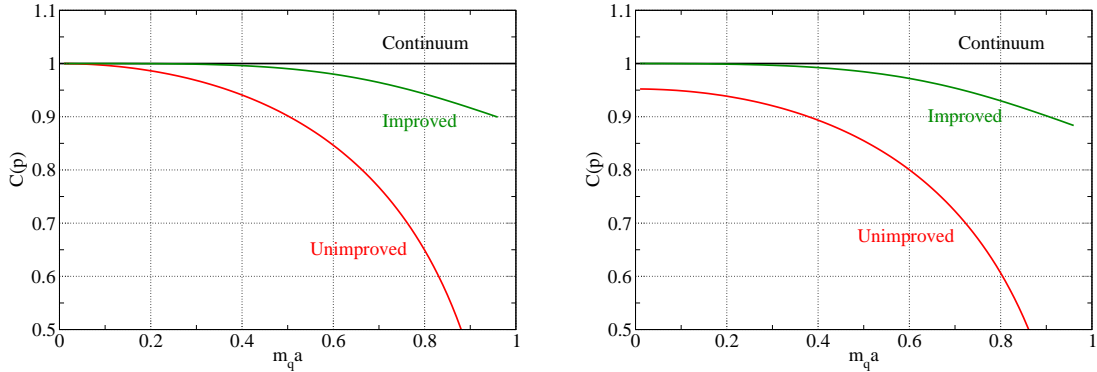

  \begin{tabular}{cc}
    \begin{minipage}{0.5\hsize}
      \includegraphics[width=0.9\textwidth]{sol_p0.eps}
    \end{minipage}
    \begin{minipage}{0.5\hsize}
      \includegraphics[width=0.9\textwidth]{sol_p1.eps}
    \end{minipage}
  \end{tabular}
  \caption{
    Effective speed of light for the ${\bf p} = (0,0,0)$ (left) and
    ${\bf p} = (2\pi/L,0,0)$ (right). The lattice volume $L = 16$ is
    assumed; $2\pi/L \simeq 0.39$.}
  \label{fig:sol}
\end{figure}

Here, we compare the improved overlap fermion action with the
unimproved one at the tree level. 
We consider the dispersion relation
\begin{equation}
  E(\vec{p}) = \sqrt{\vec{p}^2 + m_q^2} + O(a^n),
\end{equation}
which contains the lattice artifact of $O(a^n)$.
The power $n$ is 2 for the Wilson kernel while it should be 4 for the
improved kernel.
Figure~\ref{fig:dispersion} shows $E(\vec{p})$ for massless (left) and
massive (right) cases.
We can see that the improved operator certainly gives the dispersion
relation close to the continuum one.
To see more quantitatively, in Figure~\ref{fig:sol} we show the
effective speed of light defined by
\begin{equation}
  c(\vec{p})^2 = \frac{E(\vec{p})^2 - E(\vec{0})^2}{\vec{p}^2},
\end{equation}
for $\vec{p} = (0,0,0)$ (left panel) and $\vec{p} = (2\pi/L,0,0)$ at
$L$ = 16 (right panel).
The results are shown as a function of $m_q a$.
These plots imply that improved operator indeed very well reproduces
the continuum dispersion relation with only a few per cent errors up
to $m_q a\sim 0.5$,
while the unimproved operator shows much larger deviation already very
close to $m_q a=0$.

\begin{figure}[tbp]
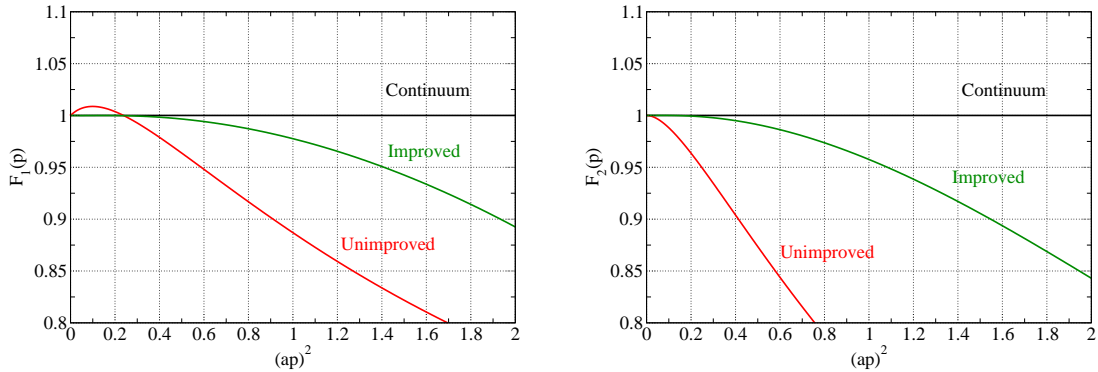

 \begin{tabular}{cc}
  \begin{minipage}{0.5\hsize}
   \includegraphics[width=0.9\textwidth]{offshell_f1.eps}
  \end{minipage}
  \begin{minipage}{0.5\hsize}
   \includegraphics[width=0.9\textwidth]{offshell_f2.eps}
  \end{minipage}
 \end{tabular}
\caption{Left panel shows $F_1(p)$ and right panel shows $F_2(p)$ versus
 $(ap)^2$. The direction of momentum is $p = (1,1,1,1)$}
\label{fig:offshell_imp}
\end{figure}

We also look at the off-shell amplitude (or the quark propagator) at
the tree level. 
We parameterize the quark propagator $S_F(p)$ as
$S_F(p) = F_1(p) \slashchar{p} + F_2(p) m_q$ after the appropriate
rotation $\Omega_L$.
We extract $F_1(p)$ and $F_2(p)$ through
\begin{eqnarray}
 F_1(p) &=& \frac{1}{4} \frac{p^2 + m_q^2}{p^2}  {\rm tr} [ i \slashchar{p}
 S_F(p)] = 1 + O(a^n) \\
 F_2(p) &=& \frac{1}{4} \frac{p^2 + m_q^2}{m^2} {\rm tr} [ m_q S_F(p)] =
 1 + O(a^n).
\end{eqnarray}
In Figure~\ref{fig:offshell_imp}, $F_1(p)$ (left panel) and
$F_2(p)$ (right panel) are shown. 
Since the improved operator has no $O(a^2)$ term, the slope of the
curve corresponding to the improved action vanishes near $(ap)^2=0$.
These plots are shown for $m_q a = 0.5$.

\section{Non-perturbative test on a dynamical lattice}
We also test the improved overlap fermion action by calculating the
meson dispersion relation.
We use the gauge configurations including 2+1 flavors of dynamical
quarks generated by the JLQCD and TWQCD collaborations
\cite{Hashimoto:2008fc}.
The lattice spacing is about $a\simeq$ 0.11~fm, and the lattice size
is $16^3\times 48$.
Sea quark masses are $m_{ud}a = 0.015$ and $m_sa = 0.080$.

For the valence quark, we use the improved overlap fermion constructed
in this work with $\rho = 1.4$.
We calculate the dispersion relation of the pseudo-scalar meson at
several different valence quark masses between 0.050 and 0.800 in the
lattice unit.

\begin{figure}[tbp]
 \vspace*{1mm}
 \begin{center}
  \includegraphics[width=8cm]{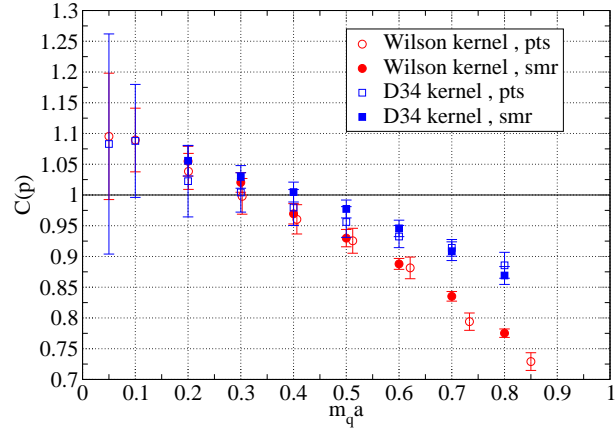}
  \end{center}
  \caption{
    Effective speed of light calculated with the improved and
    unimproved overlap fermion actions. 
    These are calculated from two smallest momentum 
    $|{\bf p}| = 0, 2\pi/L$.
    The results are shown for the overlap fermion with the Wilson kernel
 (circles) and with the improved kernel (squared).
    Open and filled symbols represent the data with a point source and
 with a smeared source, respectively.
 }
  \label{fig:sol_int}
\end{figure}

The effective speed of light is shown in Figure~\ref{fig:sol_int}.
We observe large statistical fluctuations for small valence quark
masses, as always happens for the correlators with finite momenta.
For larger quark mass region, we find that the improved operator
indeed gives the value closer to unity.
Below $m_qa\approx 0.5$, the deviation of the speed of light from 1 is
only a few per cent.

So far, we use the improved kernel as its original form.
However, the $O(a)$ and $O(a^2)$ errors in the kernel operator may
appear as radiative corrections.
We therefore should tune the parameters in the action so that these
errors vanish, which is left for future works.
Also, we are going to extend the formulation so that the improved
action produces off-shell amplitudes that are consistent with the
Ginsparg-Wilson relation. 

\vspace*{4mm}
This work is supported in part by the Grant-in-Aid of the Ministry
of Education (No. 21674002).
Numerical simulations are performed on IBM System Blue Gene Solution
at High Energy Accelerator Research Organization (KEK) under a support of its Large Scale
Simulation Program (No. 09-05).

\end{document}